\documentclass[aps,twocolumn,pre]{revtex4}
\usepackage{epsfig,amsmath,amssymb,graphicx,color}

\begin{document}
\title{Support for the value  $\text{\bf{5/2}}$ for the spin glass lower critical dimension  at zero magnetic field}
\author{Andrea Maiorano}
\affiliation{Dipartimento di Fisica, Sapienza Universit\`a
    di Roma, P.le Aldo Moro 2, 00185 Rome, Italy}
\affiliation{Insituto de
    Biocomputaci\'on y F\'isica de Sistemas Complejos (BIFI), Mariano
    Esquillor G\`omez 50018 Zaragoza, Spain}
\author{Giorgio Parisi} 
\affiliation{Dipartimento di Fisica, Sapienza Universit\`a
    di Roma, P.le Aldo Moro 2, 00185 Rome, Italy}
\affiliation{INFN, Sezione di Roma I, NANOTEC -- CNR,
    P.le A. Moro 2, 00185 Rome, Italy}

\begin{abstract}
We study numerically various  properties of the free energy barriers in the
Edwards-Anderson model of spin glasses in the low-temperature region both in
three and four spatial dimensions. In particular, we investigated the
dependence of height of free energy barriers on system size and on the
distance between the initial and final states (i.e. the overlap distance). A
related quantity is the distribution of large local fluctuations of the
overlap in large three-dimensional samples at equilibrium.
Our results for both quantities (barriers and large deviations) are in
agreement with the prediction obtained in the framework of mean field theory.
In addition, our result supports $D_{lc}=2.5$ as the lower critical dimension of the model.
\end{abstract}

\maketitle

\section{Introduction}
Many materials undergo a phase transition at sufficiently low
temperatures. It is believed that these phase transitions may be grouped into
universality classes: each universality class displays its unique behavior and
in the case of second order phase transitions, each class has its own critical exponents. Inside each
universality class, the study of the phase transitions at space dimensions
different from three  is a source of inspiration for understanding how a system behaves in our three-dimensional world. Especially in the case of
second order phase transitions it very important to get a qualitative
understanding of the properties of the system in the temperature-dimensions
plane.

Let us consider a second order transition characterized by a disordered
high-temperature phase and with an ordered low-temperature phase. Usually, there
are two special values of the space dimensions $D$:
\begin{itemize}
\item The upper critical dimension ($D_{uc}$): the critical exponents are given by the
mean field ones for space dimensions $D$ higher than the upper critical
dimension; they are non-trivial functions of the dimension $D$ below
$D_{uc}$.
\item The lower critical dimension ($D_{lc}$):  the
low-temperature phase disappears for dimensions less than $D_{lc}$. In many
cases, the transition temperature
becomes zero when we approach $D_{lc}$ and it is exactly zero at
$D=D_{lc}$.
\end{itemize}
The lower and upper critical dimensions are
universal quantities (as well as the critical exponents), in the sense that
they do not depend on the microscopic details of the Hamiltonian of the system.
We can check the soundness of our command of the physics of a model by trying
to compute these two dimensions. Failure in doing that is a symptom we miss
some crucial understanding. Indeed ignoring the upper and/or the lower
critical dimensions is actually a serious lack of understanding: in
particular, ignoring the lower critical dimensions means we lack a good
description of the mechanisms that lead to the disappearance of the low
temperature ordered phase at low dimensions.

The success of the perturbative renormalization group techniques applied to the
ferromagnetic phase transition in 3 dimensions~\cite{Wilson}, is bound to the
quantitative determination of the upper critical dimension ($D_{uc}=4$ in this case), which
in turn, allowed for the quantitative control of the infrared stable fixed
point and anomalous dimensions of operators in $4-\epsilon$
dimensions and led to the $\epsilon$ expansion for the critical
exponents.
A similar approach is suitable for obtaining useful information starting from
the knowledge of the lower
critical dimension: for example in the case of spontaneous breaking of a
continuous conventional symmetry (e.g. $O(N)$, $D_{lc}=2$) one can derive a $2+\epsilon$
expansion~\cite{2pEps}.

In the case of glassy systems, one of the theoretical difficulties is a lack of
precise knowledge about the lower critical dimension: in the case of
structural glasses, there is still a debate if a glass transition is present
in three dimensions. The best-studied case is Ising spin glasses at zero
magnetic field (i.e. the
Edwards-Anderson model). The presence of a low-temperature phase in three
dimensions has been proved experimentally and very large scale numerical
simulations do confirm the experimental result.
Franz, Parisi and Virasoro~\cite{FPV} (FPV) have done many years ago an analytic
computation of the interface free energy between different low energy
phases\footnote{An explicit computation shows that both the internal energy and the
entropy increase have a similar behavior to the one of the free
energy.}. A byproduct of this computation is the prediction that the lower
critical dimensions for spin glasses at zero magnetic field is $5/2$.
This paper aims to verify numerically the correctness of the FPV
formulae for the energy and the free energy of interfaces between different phases adding new
evidence that the lower critical dimension is $5/2$ in the case of Ising spin
glasses at zero magnetic field. Before presenting our numerical results, we will
recall the relationship between the interface free energy cost and the lower
critical dimension and recapitulate some known properties of spin glasses.

\subsection{Interfaces in ferromagnetic systems}
In the case of the Ising ferromagnet ($D_{lc}=1$) and for a Heisenberg isotropic
ferromagnet ($D_{lc}=2$), the value of the lower critical dimensions can
be computed by a simple qualitative argument based on the cost of the free energy
for creating an interface between two regions with different values
of the order parameter.
There is general consensus on impossibility of long-range order when such
cost is finite in the thermodynamic limit.
Sometimes the computation can be simplified by computing the
increase in the ground state energy at zero temperature upon changing the
boundary conditions in an appropriate way.

Let us see how to do such a computation in the Ising
ferromagnet:
The spins are $\pm 1$ variables. We consider
a $D$ dimensional system, with periodic conditions in all directions $x_2,\dots,x_D$,
but in the $x_1\equiv x$ direction where we impose fixed boundary conditions. In the
plane $x=0$ we set $\sigma=1$ and in the plane  $x=L-1$ we set either
$\sigma=1$ (periodic boundary conditions) or $\sigma=-1 $ (antiperiodic boundary conditions). Our aim is to
compute the ground state energy difference as a function of $L$. In the case of
periodic boundary conditions the ground state is  all $\sigma$'s equal to 1,
while in the case of antiperiodic boundary conditions the ground state is
given by $\sigma(\vec{x})=1$ for $x_i<M, \; 0<M<L-1$ and $\sigma(\vec{x})=-1$ for $x_i\ge M$. We immediately get
the variation $\Delta E(L)$ of the
energy is $2L^{D-1}$.

If we are interested in computing the free energy difference at non-zero
temperature, not too near to the critical point, we can write a
Landau-Ginzburg-like expression for the magnetization $m(\vec{x})$. One finds
that the variation of the free energy is $\Sigma(T)L^{D-1}$, where $\Sigma(T)$
is the surface tension. The free energy increase of the free energy in $D=1$
goes to a constant for large $L$ and therefore $D_{lc}=1$ is the lower
critical dimensions.

In the planar spin model case  spin waves are present and we can have smooth
interfaces with much lower free energy cost. In this model the spins are two
dimensional vectors of modulus 1 and they can be parametrized as
$\sigma(\vec{x})=\{\cos\left(\theta(\vec{x})\right),\sin\left(\theta(\vec{x})\right)\}$. Neglecting
vortices, in the continuum limit  the phase $\theta(\vec{x}) $ is a smooth
function.
In the low temperature regime we can write an effective free energy as
\begin{align}
AL^{D-1}\int d^D x\left(\frac{d \theta(\vec{x})}{dx}\right)^2\,.
\end{align}
where  $\theta(\vec{x})$ denote the direction of the magnetization around the
point $\vec{x}$ and  $A$ is a positive constant. This expression can be
derived from a Landau Ginsburg functional  (or  equivalently from the
Goldstone model) where one neglects the longitudinal fluctuations in the
direction of the magnetizations.

In this case we can introduce more complex boundary condition:
e.g. $\theta(x_1)|_{x_1=0}=0$ and $\theta(x_1)|_{x_1=L-1}=\theta_B$. A detailed
computation (see appendix~\ref{app:BarriersH}) tells us in this case we can construct an interface where
the  phase $\theta(\vec{x}) $ is a smooth function.  We find that the free
energy  increases as $A L^{D-2} \theta_B^2$, hence $D_{lc}=2$. A similar
results is obtained for the internal energy. Indeed in
$D=2$ these differences remain of order 1, also when
$L\to\infty$.

The absence of a phase with a non-zero order parameter in two
dimensional systems is the essence of the Mermin-Wagner-Hohenberg
\cite{MerminWagner, Hohenberg} theorem where one studies small fluctuations around
equilibrium, proving that in presence of non-zero order parameter the
correlation function in the small momentum region behaves as $1/k^2$ (i.e. a
Goldstone Boson is present) and this behavior is inconsistent in a
2D world.

\subsection{Spin glasses: experimental and numerical results}

A popular model of spin glasses at zero magnetic field is the
\emph{Edwards-Anderson} model~\cite{EA} (EA) in which the Ising spins
$\{\sigma\}$ are arranged on a $D$ dimensional cubic lattice. Only
interactions among nearest neighbors pairs contribute to the energy: the
Hamiltonian is given by
\begin{align}
\label{eq:H}
H=-\sum_{\langle i,k\rangle}J_{i,k}\sigma_i,\sigma_k\;,
\end{align}
where the $J_{i,k}=\pm1$ are quenched (frozen) random couplings. Different
realizations of the configuration of couplings $\{J\}$ define different
\emph{instances} (or \emph{samples}) of the system.
Two or more independent copies of the same instance are called \emph{replicas}.

In the low-temperature phase a crucial quantity that plays the role of order
parameter is the expectation value of the overlap $q_i\equiv\sigma_i\tau_i$,
where $\sigma_i$ and $\tau_i$ are spins at site $i$ in any two independent equilibrium
configurations. We define the {\sl intensive} value of the overlap in a box of linear size $L$ as
\begin{align}
\label{eq:Q}
Q=\frac{1}{L^D}\sum_i q_i\; .
\end{align}
In the mean field approximation the thermal average $\langle
Q^2\rangle_J$ for a given disorder instance is non-zero below the transition
temperature in the infinite volume limit. For a given sample the overlap may
take many different values and with changes in the {\em extensive} free
energy that are of order 1: in
other words there are globally different  arrangements  of the spins that have
comparable probabilities. As a consequence the overlap probability distribution function
$P_J(Q)$ (see details in the appendix~\ref{app:Q}) is of order 1 for many different values of
$Q$: $P_J(Q)$ depends on the choice of
the $J$'s (non-self averageness); also $\langle Q^2\rangle_J$   depends on the
values of the $J$'s.  If we take two different equilibrium
configurations of the system (e.g. $\{\sigma\}$ and $\{\tau\}$) their global overlap
$Q$ can be in the range $[-q_{EA}:q_{EA}]$, where, denoting by
$\left[\cdots\right]_J$ the average over samples,
$q_{EA}=\left[\langle s_i\rangle^2\right]_J$ is the so-called \emph{EA order
parameter}, without any additional cost. The
$Q$-constrained free energy density $F(Q)$ is constant in this interval as
shown by an explicit computation. The existence of flat regions in
free energy has deep consequences.

The mean field theory is relatively well understood: it is valid in the simple
Sherrington Kirkpatrick model that naively correspond to the
infinite-dimensional limit of the Edwards-Anderson model.
In finite dimensions, the analytic studies are
more complex. Standard arguments can be used to construct a low momentum
effective Landau-Ginzburg theory and apply renormalization group-like
techniques~\cite{LGWeff}.
The system has a standard second-order phase transition with a
divergent nonlinear susceptibility $\chi^{\text{(3)}}(T)\propto (T-T_c)^{-\gamma}$. It
has been shown that in dimensions greater than $6$ the critical exponents are
those of mean field (i.e. $\gamma=1$). An $\epsilon$ expansion for the
critical exponents has been constructed in $\epsilon=6-D$~\cite{HLCcrit}: the series have been
computed up to the order $\epsilon^3$~\cite{Green}  but unfortunately the convergence of the
series is not good and it is difficult to use them already in dimensions $D=5$.

Quite accurate experiments~\cite{TcExp} and numerical
simulations~\cite{TcNum} agree on the existence
of a transition in dimensions $D=3$, with quite a large value of $\gamma$
(i.e. $\gamma\approx 6$), therefore $D_{lc}<3$. On the contrary in dimensions
$D=2$ the non-linear susceptibility is finite at any positive temperature and it
has a power-law divergence in the zero temperature limit. According to standard
folklore, at the lower critical dimensions, the relevant susceptibility should
diverge exponentially when the temperature goes to zero and a power-law divergence
should be present only below the lower critical dimensions. Numerical
simulations done by Boettcher~\cite{Boettcher1,Boettcher2} gives
$D_{lc}=2.4986$ with a small error.
Boettcher studies the exponent that controls the
dependence on $L$ of the variance of the energy difference from periodic to
antiperiodic boundary conditions for different dimensions $D$.  This exponent
should change sign at the lower critical dimension: its value is obtained by
interpolating (as a function of the dimensions $D$) this critical exponent. A
similar estimate $D_{lc}=2.491$ (not accurate as the previous one) comes from
the extrapolation of the values of the critical temperature~\cite{Boettcher3}
as a function of dimensions, assuming that $T_c$ vanishes proportionally to $\sqrt{D-D_{lc}}$, as
suggested by theoretical considerations~\cite{FH} \footnote{A less
quantitative prediction comes from the value of the exponent $\eta$ as a
function of the dimensions $D$. This exponent strongly decreases when going
from dimensions $4$ to dimensions $3$ where its value is $\eta(3)=-0.39\pm
0.1$~\cite{Jcrit}.  We expect that at the critical dimensions $D_{lc}=2-\eta(D_{lc})$.  A
simple extrapolation of  $\eta(D)$ would give $D_{lc}\approx 2.6$ with large
errors. If we assume that $\eta(D)$ is a decreasing function of $D$ (as it
happens in all the extrapolations) the three-dimensional value implies that
$D_{lc}>2.4$.
}.

Summarizing, numerical simulations and experiments tell us that
$2<D_{lc}<3$. There are also strong numerical pieces of evidence that the value of
$D_{lc}$ is quite near to and likely equals $5/2$ \footnote{We expect that the
upper and the lower critical dimensions are simple rational numbers in the rare case that they are not integers (non-integer values are possible: for example the upper critical dimension for a quadricritical point - that is represented by a $\phi^8$ interaction - is $8/3$).}.

\begin{figure}
\centering
\includegraphics[width=0.9\linewidth]{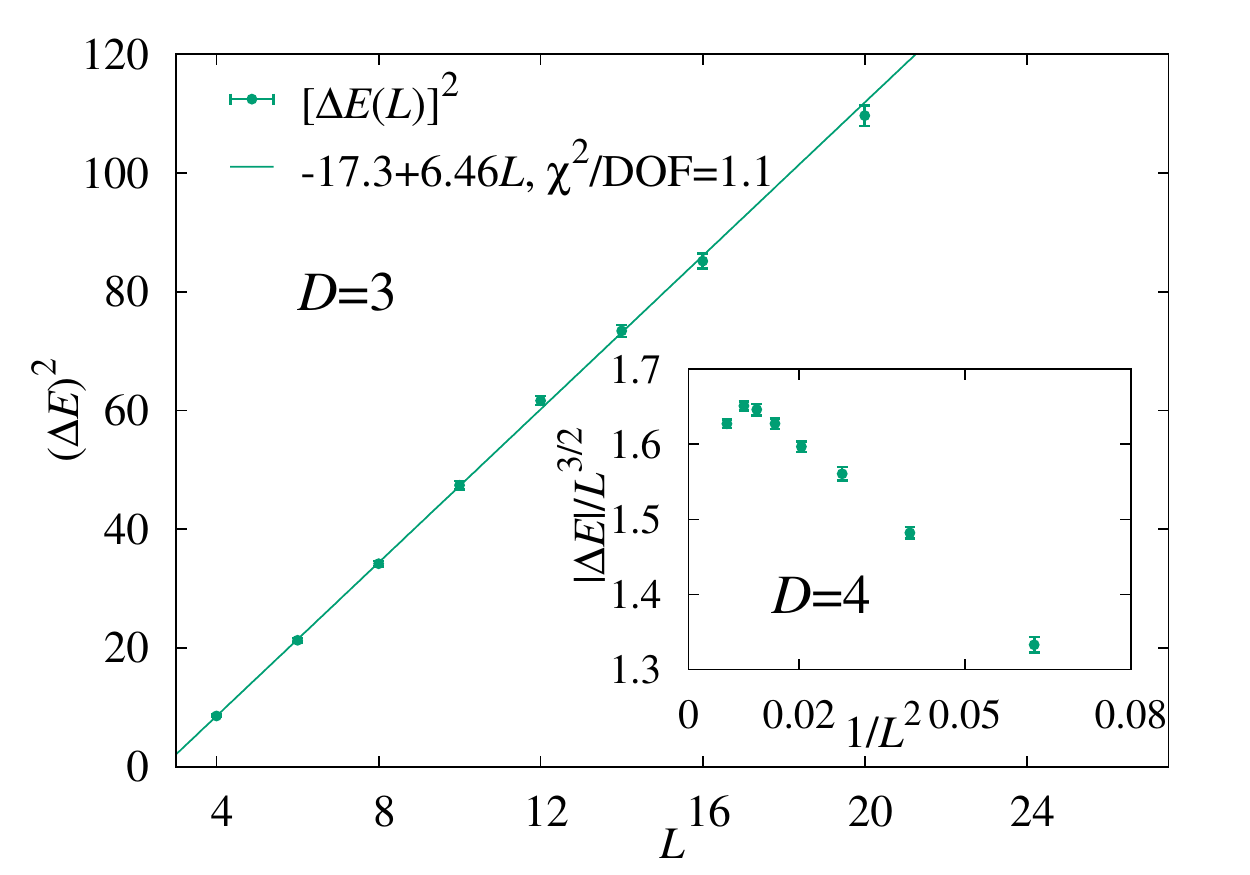}
\caption{Main plot, $D=3$: $\Delta E(L)^2$ as a function of $L$ at $T=0.7$. The theory
predicts an asymptotic linear behavior. In the inset, $D=4$:  $\Delta
E(L)/L^{3/2}$ as function of $1/L^2$ at $T=1.4$. The theory predicts a finite non zero
limit at $1/L=0$.}
\label{fig:DE}
\end{figure}

\subsection{Interfaces in spin glasses}

The value $D_{lc}=5/2$ was predicted in 1993, much before Boettcher's work, in
a remarkable paper \cite{FPV} assuming that there are  minimal corrections to
mean field theory predictions\footnote{It was assumed that we have a
Landau-Ginsburg type functional whose form is obtained neglecting loop corrections to mean field theory.}.

We have seen that we can define a free energy as a function of the global
overlap $Q$ and the $Q$-constrained free energy density $F(Q)$ is constant for
$|Q|<q_{EA}$. A natural question arises when we constrain two different large
regions of the system to have two different values of $Q$. The bulk
contribution to the free energy vanishes, and we remain with the contribution
coming from the interface that we have to evaluate.

In the paper \cite{FPV}, FPV considered   two  systems A and B with the same
Hamiltonian (i.e. same $\{J\}$) inside a $D$ dimensional box of side
$L$. Using the same geometry that we have discussed above,  they studied  the
free energy increase when we   constrain the two systems to
have a mutual overlap $Q^{AB}$ with a value
$Q^{AB}=Q$ on a plane on the boundary at $x=0$ and  $Q^{AB}=Q'=Q+\Delta$ at
the other boundary at $x=L$. The computation was done for small $\Delta$ in
the region where both $Q$ and $Q'$ are in the range $[-q_{EA}:q_{EA}]$. We can write
$q(x)=Q+\theta(x)$ with
\begin{align}\theta(0)=0 \quad \mbox{and} \quad\theta(L)=\Delta\,. \label{eqBC}
\end{align}
One can thus compute the free energy cost by using a variational procedure with
respect to all other variables (i.e. probability distribution of all the
overlaps $Q^{AA}$ for system A, $Q^{BB}$ for system B and all the overlaps
$Q^{AB}$ except those on the boundary).  One finally arrives at the expression for the free energy increase
$\Delta F_L[\theta]$ that is a functional of $\theta(x)$. At the end of the day, we have to minimize $\Delta F_L[\theta]$.

The results of this explicit computation
were rather surprising.
A simple quadratic analysis of the free energy \footnote{We keep the
quadratic terms in $\theta$ in the free energy or equivalently the linear ones in the mean-field
equations.} implies that the final result is the sum of a few terms of the form
\begin{align}
\Delta F_L[\theta]=AL^{D-1}\int_0^L d x\left(\frac{d \theta(x)}{dx}\right)^2\;.
\end{align}
However when all the terms are assembled these quadratic contributions cancel
out. No free energy increase is present if we consider only quadratic
terms.

If we keep higher-order terms (e.g. the cubic terms) we get nonlinear terms
in the mean field equations. Finally we obtain the amazing
result:
\begin{align}
\label{eq:Ftheta}
\Delta F_L[\theta] \propto  L^{D-1}\int_0^L d x\left(\frac{d\theta(x)}{dx}\right)^{5/2}\,.
\end{align}
This result can also be generalized to  the case of a function $\theta(\vec{x})$ that depends on all the coordinates of $\vec{x}$.
\begin{align}
\Delta F_L[\theta] \propto \int_0^L d x^D \left(\sum_{\nu=1,D}\left(\frac{\partial \theta(\vec{x})}{\partial  x_\nu}\right)^ 2\right)^{5/4}\,.
\end{align}
We finally obtain that the free energy increases is given by
\begin{align}
\label{eq:FPV}
\Delta F(L,\Delta)\equiv\min_{\theta} \Delta F_L[\theta]\propto   L^{D-5/2} \Delta^{5/2}\; ,
\end{align}
where the minimum is done over all the functions $\theta(x)$ that satisfy the boundary conditions eq. (\ref{eqBC}). A similar expression is obtained for the internal energy.

This analysis implies that the barriers are much smaller than in the known
cases of spontaneous breaking of a continuous symmetry (the nature of the
Goldstone modes is quite different). At the end we find that the barriers
vanish for $D\le 2.5$, hence $D_{lc}=2.5$.

\section{Results and discussion}

It is clear that the validity of the FPV result, which has been derived in the
mean field framework, can be considered doubtful. However, a similar result
also holds in the ferromagnetic case, when other properties of mean field
theory are not valid. Indeed detailed arguments show that the interaction of Goldstone Bosons (magnons in this
case) is essentially the same as in mean field theory.

The FPV theory predicts a value of the lower critical dimension that
is very near to the one suggested by earlier numerical simulations. In order to check
its validity beyond the assumptions used for its derivation, we
have investigated the agreement of its predictions with the results of
purposely designed numerical simulations. As we shall see the results are in remarkable
agreement.

We will consider two different kinds of simulations: the direct measurement of
the interface energy and the study of large deviations of the overlap
differences in the same sample at equilibrium.

\subsection{Direct measurement of the Interface Energy}

We have computed directly the interface energy $\Delta E(L,Q,\Delta)$ in
$D=3$ and $D=4$ in the most extreme case $Q=0$ and $\Delta=2$, i.e. $Q=1$
on one boundary and $Q=-1$ on the other boundary.  We have studied this
extreme case for two reasons: the signal to noise ratio is higher and its
numerical implementation is much simpler that than in the case where
$\Delta<2$.

An FPV computation predicts that both the interface energy $\Delta E(L)$ and
interface free energy $\Delta F(L)$ grow as $L^{1/2}$ for $D=3$ and as
$L^{3/2}$ for $D=4$. We have studied only the internal energy that can be
computed in a much simpler way than the free energy.

The data we will discuss below have been produced by simulations of the
EA model equation~\ref{eq:H} with binary couplings ($J_{ij}=\pm1$ with equal
probability) in $D=3$ and $D=4$. (see section \emph{Methods} below for details). The critical temperature for the model
is $T_c\simeq1.103$ in $D=3$~\cite{Jcrit} and $T\simeq 2.0$ in $D=4$.~\cite{D4Tcrit}
We have done our simulations at a temperature of the order of $0.7 T_c$.
At this value of the temperature thermalization of the
samples is not too difficult, and we are far enough from the critical
temperature for simulations to be not too sensitive to crossover effects. More precisely our
simulations have been done at $T=0.7\simeq0.64T_c$  in $D=3$ and at
$T=1.4\simeq 0.7T_c$ in $D=4$.

We report data for the square of the interface energy $\Delta E(L)^2$ as a
function of the linear size up to $L$ for both $D=3$ (up to $L=20$) and $D=4$
(up to $L=12$) in Figure \ref{fig:DE}.

A linear growth of $\Delta E(L)^2$ describes very well the $D=3$ data, in very good agreement
with the theoretical prediction that gives $\Delta E(L)\propto L^{1/2}$. A
linear fit to $\Delta E(L)$ works surprisingly well also at $L$ as small as 4
up to the largest value of $L$, i.e. $L=20$.  The reasons for such small
finite size corrections in $D=3$ are unclear.

In dimensions $D=4$ data have stronger finite size
corrections, but the ratio $|\Delta E(L)|/L^{3/2}$ tends to saturate at larger
sizes. Unfortunately, we are limited to consider values of $L$ up to 12 in
$D=4$ - that correspond to 20736 spins, a number already much larger than the number of spins (8000) that we used in $L=20$ for $D=3$.

Both three-dimensional and four-dimensional data strongly support the
prediction $\Delta E(L)\propto L^{D-5/2}$.

The results we obtain for the scaling exponent of energy
differences are larger than previous numerical estimates by measuring the energy cost of
flipping boundary conditions in ground state computations~\cite{Hartmann,
Boettcher3}. We stress we have a completely different setting here, the main difference being the imposed constraint
(fixed total overlap $Q$ and overlap difference between opposite boundaries
$\Delta$) determining completely different excitations (see discussion in the appendix~\ref{app:cfr}).

\begin{figure}[t]
\centering
\includegraphics[width=0.9\linewidth]{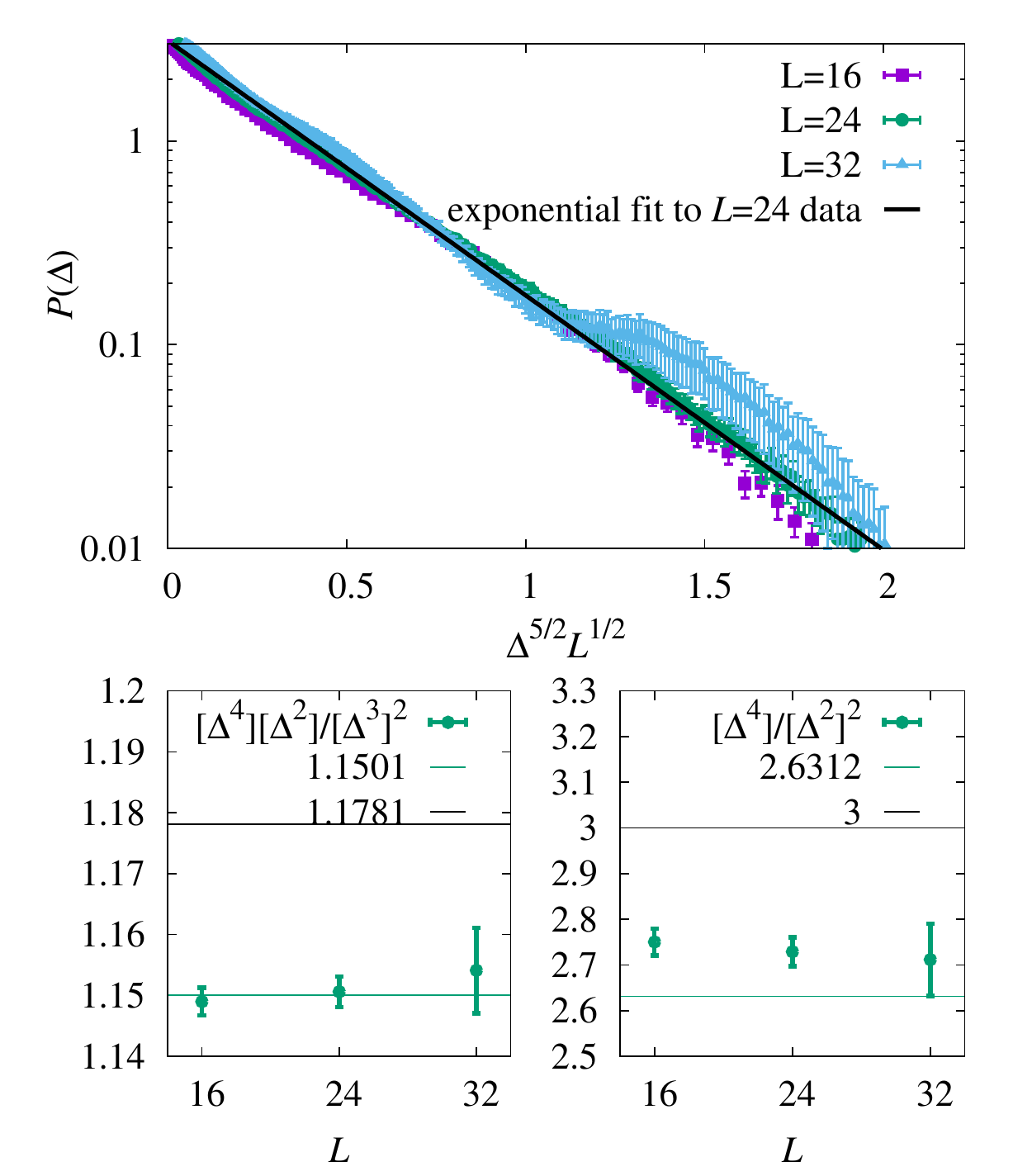}
\caption{Top panel: $P(\Delta_M,L)$, $M/L=1/8$ (see text) as function of $z\equiv
L^{1/2}\Delta^{5/2}$, $T\simeq0.64T_c$. Bottom panels: Comparison of cumulants obtained from
the numerical data (points with errorbars) and values obtained with the prediction
equation~\ref{eq:Pdelta} (green line)
and expected for a Gaussian distribution (black line); right: the Kurtosis
$K$; left: the cumulant ratio
$R$ (see text).}
\label{fig:Pdelta}
\end{figure}

\subsection{Large deviations of local overlaps fluctuations in a sample}
We could use the previous approach to study also the $\Delta$
dependence of the Energy barriers by
performing a different simulation for each value of $\Delta$. Here we prefer to do the direct tests
of equation \ref{eq:FPV} for the {\sl free} energy,  computing the probability of
rare configurations in existing large scale simulations performed by the Janus collaboration~\cite{JanusEq, Janus}.

In the low-temperature phase, the local overlaps $q(\vec{x})$ of two
equilibrium configurations should fluctuate around the volume average of $q$,
i.e. $Q$. The probability of having a rare fluctuation with an overlap value $q$
significantly different from $Q$ in a large region is exponentially damped
and it can be computed starting from equation (\ref{eq:Ftheta}). The
computation could be done by means of a standard simulation of spin glasses and
looking for the probability distribution of these rare events. A computation
along this baseline for hierarchical spin glass models on Dyson lattices can
be found in~\cite{FJPhierarchical}. We find
convenient to use the large database of the Janus collaboration that contains
already thermalized spin glass configurations for quite large lattices (up to
$L=32$ and down to $T=0.64T_c$ in $D=3$).

The region where we look for large fluctuations of $Q$ may be a cube, as in
the case of window overlaps, however in this case we consider a very simple
geometric setting.
Let us define the quantities of interest. We work in $D=3$ in a box of size
$L$ with periodic boundary conditions. We define the  overlap  $q_{M,x}$,
obtained by averaging the local overlap in a region of size $L^2\times M$
delimited by $x\le i_x \le x+M-1$:
\begin{align}
q_{M,x}=\frac{1}{ML^2}\sum_{x\le i_x \le x+M-1}\sum_{i_y,i_z} q(i_x,i_y,i_z)
\end{align}
We are interested in computing the probability of having $q_{M,x}$ quite
different from the global average $Q$.
In order to simplify the analysis we define $\Delta_M=\frac{1}{2}|q_{M,x}-q_{M,x+L/2}|$,
i.e. the difference in the overlap of two regions of size $L^2\times M$ that are
at the largest possible distance (we are using periodic boundary
conditions), normalized in $[0,1]$

The quantity of interest is the probability density of $\Delta_M$ inside a box of size
$L$, i.e. $P_M(\Delta_M,L)$, with fixed total overlap in the two regions
$Q_M=\left|q_{M,x}+q_{M,x+L/2}\right|$, in the large deviation region where this
probability is small; we consider $Q_M=0$ (see section \emph{Methods})
allowing for the largest range of fluctuations $\Delta$ and more statistics in
the large deviation region.
We average $P_M(\Delta_M,L)$ over all samples. In the large deviation region, this probability is given by the exponential of  the free energy difference multiplied by $-\beta$.
The prediction of FPV is
\begin{align}
\label{eq:Pdelta}
P_M(\Delta_M,L) \propto \exp\left(-A_{M,L} L^{1/2}\Delta_M^{5/2}\right)
\end{align}
in the large deviation region $z\equiv L^{1/2}\Delta_M^{5/2}>> 1$ and $\Delta_M$ not too large.
The coefficient $A_{M,L}$ does depend on the details of the free energy and
therefore it cannot be computed: however, we can compute its dependence on
$M$: as we shall see it turns out to be a function of $M/L$.

We plot  $P_M(\Delta_M,L)$ in $D=3$  at $T=0.7$ as a function of
$z=L^{1/2}\Delta^{5/2}$ for $L=16,24,32$ and $M/L=1/8$ in
Figure~\ref{fig:Pdelta}, top panel. The $L=32$ data at the largest z
values are noisy, due to the lower statics that we have at this value of $L$.
The theoretical  prediction $\exp (-Az)$  is accurate in almost all the range,
showing deviations from an exponential decay only at very small $z$ values: these
deviations are an expected effect because at small $\Delta_M$  we must have
$P_M(\Delta_M,L)=P_M(0,L)-O(\Delta_M^2)$.

We have computed the cumulant ratios
\begin{align}
K=\langle\Delta^4\rangle/\langle\Delta^2\rangle ^2\,, \label{eq:cum} \qquad
R=\langle\Delta^4\rangle \langle\Delta^2\rangle/\langle\Delta^3\rangle^2
\end{align}
whose numerical values (depicted in Figure~\ref{fig:Pdelta}, bottom panels)
compare well with the values predicted using equation~\ref{eq:Pdelta}. These are
only approximate predictions because both $K$ and $R$ depend on the behavior
of $P_M(\Delta_M,L)$ in the region of small $z$ where the large deviation
behavior $\exp(-A z)$ is not expected to hold. The
ratio $R$ has been constructed in such a way to be less dependent on the
value of the probability in the small $z$ region: its value is remarkably in
better agreement with the theoretical predictions than the Kurtosis $K$.

\begin{figure}[t]
\centering
\includegraphics[width=0.9\linewidth]{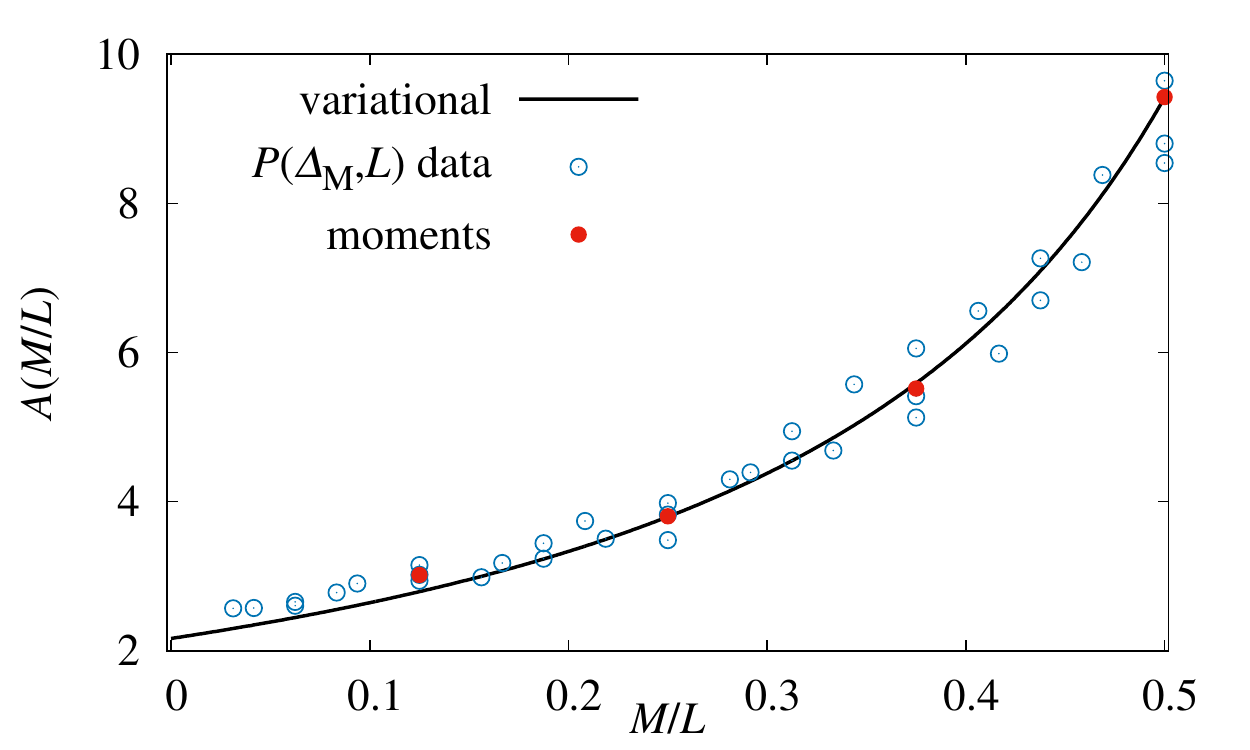}
\caption{The coefficient $A_{M,L}$ in the energy barrier (see
equations~\ref{eq:Pdelta2},~\ref{eq:moms}), obtained by i) fitting Monte Carlo $P(\Delta_M;L)$
data (open circles) with $M=1,\dots,L/2$ and $L=16,24,32$; ii) fitting the formula equation~\ref{eq:moms} to the second
moment $\langle \Delta^2 \rangle$ (bullets) at ratios $M/L=1/8,1/4,3/8\mbox{
  and } 1/2$; iii) variational computation in
the continuum limit - see appendix~\ref{app:var} (continuous line). An overall factor to the variational
data has been adjusted to match moments data (see text).}
\label{fig:A}
\end{figure}

We have also looked at the dependence of $A_{M,L}$ as a function of $M/L$. The
theoretical predictions (derived in the \emph{Methods} section) are shown in
fig. \ref{fig:A} and are in very good agreement with the numerical data.

\section{Conclusions} The numerical evidence presented above strongly support the correctness of the FPV
prediction on the lower critical dimension $D_{lc}=5/2$ and the scaling of the
free energy interface barriers. It is remarkable that the corresponding
exponents have a simple functional dependence on $D$  and they do not show any
of the usual anomalous corrections when extending below the upper critical
dimension. This phenomenon is typical of interface energies in the broken
symmetry phase where these quantities are not changed by corrections to mean
field theory: it is also related to the decoupling of Goldstone type modes at
low momenta. Here the situation is far more complex because the analysis is
based on non-linear corrections and it does not match with perturbative
corrections. 

In spin glasses, one can define constrained connected correlation functions:
$C(x|Q)\equiv \langle q(x) q(0)\rangle_Q-Q^2$, where the average is done in a
two replica system where the total overlap is $Q$. Dimensional
analysis implies that $C(x|Q)= B(Q) x^{-\alpha}$ with $\alpha=4/5(D-5/2)$ in the
region of $Q<Q_{EQ}$.
In $D=3$ {\sl large} lattices simulations give $\alpha(Q)$ independent
from $Q$, equal to $0.38\pm 0.02$ against the theoretical prediction of
$2/5$ \cite{Jtheta}.

In momentum space the FPV prediction becomes
$
\tilde C(k|Q)\propto 1/(k^2)^{D/5+2}\; .
$
%D-4=4/5D-2; 1/5D=2  ;D=10
%D-3=4/5D-2; 1/5D=1
%\vskip50pt
%
This last prediction on the momentum behavior poses more questions. An
expansion around mean field~\cite{DDKT} in high dimensions gives
two different exponents $\alpha$ depending on the value of
$Q$. At $Q=0$: $\alpha(0)=D-4$;
at $0<Q<Q_{EA}$: $\alpha(Q)=D-3$.
These two predictions cross the FPV prediction at $D=10$ for $Q=0$ and at
$D=5$ for $Q\ne 0$. It is
unclear if something special happens at these two dimensions.

\section{Methods}

\subsection{Direct measurement of the Interface Energy}

We performed Monte Carlo simulations of the Edwards-Anderson model with binary
couplings, equation~\ref{eq:H}, on the cubic lattice of size $L$ with periodic
boundary condition for
$L=4,6,8,10,12,14,16,20$ in $D=3$ and $L=4,5,6,7,8,9,10,12$ in $D=4$ by means
of a single spin flip dynamics with the usual Metropolis
algorithm and using Parallel Tempering~\cite{PT} to improve decorrelation and convergence.
The lowest temperatures simulated in the Parallel Tempering protocol are
$T=0.7\simeq0.64T_c$ for $D=3$ and $T=1.4\simeq0.7T_c$ for $D=4$ which are
also the temperatures for which we show data in
this work. We simulated $N_J=12800$ different instances of the system.

The simulation protocol we used to measure the interface barrier in the
large $\Delta$, $Q=0$ sector is the following:

1. We thermalize a given instance of the system $S$;

2. Once equilibrium is reached, the system is replicated twice: the replica $S^{(+)}$ retains periodic
boundary condition; let's label site coordinates as $i={i_1,\dots,i_{D}}$ in $D$
dimensions; we apply antiperiodic boundary conditions in the $D$-th
direction to the replica $S^{(-)}$;

3. We freeze spins on the $i_D=0$ (hyper-)plane on both $S^{(+)}$ and
$S^{(-)}$ (we inhibit their update in the single spin flip dynamics);

4.  We thermalize both $S^{(+)}$ and $S^{(-)}$ and
compute $\Delta E = \langle E(S^{(-)})-E(S^{(+)})\rangle$
where $\langle \cdots \rangle$ is a
thermal (Monte Carlo) average.

5. We repeat and collect statistics of $\Delta E$ over $N_J$ samples.

If we compute overlaps between $S^{(+)}$ and $S^{(-)}$ on planes at fixed
$i_D$
\begin{align}
\label{eq:qx}
q(x)=\frac{1}{L^{D-1}}\sum_{i_1,\dots,i_{D-1}} \sigma_{i_D=x}^{(-)}\sigma_{i_D=x}^{(+)}\; ,
\end{align}
in the limit of large $L$, $\Delta=|q(x=0)-q(x=L-1)|$ tends to the maximum
allowed value, while $L^{-1}\sum_xq(x)\rightarrow 0$.

A related approach has been followed before in~\cite{CGGPV2011}
to study the scaling properties of the interface energy in a different
setting and with smaller system sizes, obtaining compatible results.

\subsection{Large deviations of the overlap fluctuations}
To extract data for the distribution $P(\Delta_M, L)$ we took profit of the
Janus Collaboration's database~\cite{JanusEq, Janus}. The dataset consists of many
equilibrium configurations at different temperatures (in the number of
O(100) independent spin configurations per sample and per
temperature at the largest size) of 4000 samples of size $L=16,24$ and 1000
samples of size $L=32$ of the $D=3$ Edwards-Anderson model with binary couplings.

For each pair of spin configurations $\{\sigma\}$ and $\{\sigma^\prime\}$ we
compute the
overlaps in boxes of size $ML^2$:
\begin{align}
\label{eq:qMz}
q_M(z)=\frac{1}{ML^2} \sum_{z\le i_D\le z+M-1}\sigma_i \sigma_i^\prime \; ,
\end{align}
for $M=1,\dots,L/2$ and collect statistics for
$\Delta_M=\frac{1}{2}|q_M(z)-q_M(z+L/2)|$
in the sector $Q_M=\frac{1}{2}\left|q_M(z)+q_M(z+L/2)\right|<1/16$ (this is an
arbitrary cutoff chosen to soften the $Q_M=0$ constraint enough to have satisfactory
statistics; other $1/2$ factors are chosen to normalize $Q_M$ and $\Delta_M$ in
$[0,1]$). The FPV prediction for the distribution of $\Delta_M$ is
\begin{align}
\label{eq:Pdelta2}
P(\Delta_M,L) \propto \exp\left[ -A_{M,L} L^{1/2}\Delta_M^{5/2}\right]\; ,
\end{align}
where the constant $A_{M,L}$ depends on the definition of the
overlap, mainly on the boxes geometry through the ratio $M/L$. The moments of
the distribution in equation~\ref{eq:Pdelta2} are combinations of Euler's gamma functions:
\begin{align}
\label{eq:moms}
\langle \Delta_M^k \rangle =
A_{M,L}^{2k/5}L^{-k/5}\frac{\Gamma\left((2k+2)/5\right)}{\Gamma\left(2/5\right)}\; ,
\end{align}
and cumulants such as those in equation~\ref{eq:cum}, should not depend on
either $A$ or $L$.
For large $L$ the coefficients $A_{M,L}$ should not depend on $L$ as long as
$M/L$ is kept fixed. We estimate values of $A$ from our data in two ways: i)
by fitting equation~\ref{eq:Pdelta2} to $P(\Delta_M,L)$ data; ii) by fitting
equation~\ref{eq:moms} to $L$ dependent data at fixed $M/L$. Results are shown in
figure~\ref{fig:A}. The extracted values compare well to estimates obtained
in independent computations in the continuum limit (see appendix~\ref{app:var}),
which are represented in figure~\ref{fig:A} as a continuous line.

\acknowledgments{
This project has received funding from the European Research Council (ERC)
under the European Union’s Horizon 2020 programme (grant No.694925).
We thank S. Franz, who stimulated us in checking the FPV predictions on the scaling exponents, 
V. Astuti for very useful discussions on $\Delta F(L,\Delta)$, 
and the Janus collaboration for letting us analyze Janus data.
}

\appendix

\section{Barriers in the Heisenberg ferromagnet}
\label{app:BarriersH}
Let us see how to do such a computation in the Heisenberg
ferromagnet: the spins are unit vectors on a three-dimensional sphere \cite{Dr, SFT}. We consider
a $D$ dimensional system, with periodic conditions in all directions $x_2,\dots,x_D$,
but in the $x_1\equiv x$ direction where we impose fixed boundary conditions. In the
plane $x=0$ and $x=L-1$ we set  respectively
$\sigma=\{1,0,0\}$ and $\sigma=\{\cos(\theta_B),\sin(\theta_B),0\}$. Our aim is to
compute the ground state energy as a function of $L$ and $\theta_B$. For
convenience, we consider only the variation $\Delta E(\theta_B,L)$ of the
energy with respect to the ground state energy with periodic boundary
conditions.

To this end we can  consider only spins of the form
$\sigma(\vec{x})=\{\cos\left(\theta(\vec{x})\right),\sin\left(\theta(\vec{x})\right),0\}$. In the
limit of large $L$ the space dependent phase $\theta(\vec{x}) $ is a smooth function; neglecting lattice effects we can write
\begin{align}
\Delta E(\theta_B,L)=A\int_0^L d x^D \sum_{\nu=1,D}\left(\frac{\partial \theta(\vec{x})}{\partial  x_\nu}\right)^2\,,
\end{align}
where $A$ is a positive constant.

Translational invariance is esplicitely broken by the boundary conditions
only in the $x$ direction, so it is safe to assume that $\theta(\vec{x})$ is
a function of only $x_1\equiv x$. We thus arrive to
\begin{align}
\Delta E(\theta_B,L)=
AL^{D-1}\int_0^L d x\left(\frac{d \theta(x)}{dx}\right)^2\,.
\end{align}

If $\theta_B=0$, i.e. periodic boundary conditions, the ground state is
obviously given by $\theta(x)=0$: $\Delta E(0,L)=0$. If $\theta_B\ne 0$, the energy
minimum is given by a smooth interface $\theta(x)=\theta_Bx/L$.
The  energy increase is   $A L^{D-2} \theta_B^2$, hence $D_{lc}=2$. Indeed in
$D=2$ the energy difference remains of order 1, also when
$L\to\infty$. Showing that the same argument can be used also at non zero
temperature to prove the nonexistence of ordered phases requires a more
complex proof.

The proof of the Mermin-Wagner-Hohenberg theorem
\cite{MerminWagner, Hohenberg} about the absence of a phase with non-zero
order parameter in two
dimensional systems is quite different from the
one presented here. In the MWH proof, one considers small fluctuations around
equilibrium, here one considers large deviations from equilibrium, however,
the physics foundations are the same.

If we extend this computation to the free energy at finite temperature, we
obtain similar results, and the same method can be used to study the behavior of
the correlations functions. We can get the final result in a fast way if we use
dimensional analysis. Indeed, with an appropriate rescaling, we can set $A=1$
and the combination $L^{D-2} \theta_B^2$ becomes dimensionless. If we assign
to the length dimension -1, $\theta(x)$ must have dimensions $(D-2)/2$. Dimensional
counting implies then that the correlation $C(x)\equiv \langle \theta(x)\theta(0)\rangle$ decays as
$|x|^{-\alpha}$ with $\alpha= D-2$. The
same formula in momentum space reads as $\tilde C(k)\propto 1/k^2$, that is
what we expect from the Goldstone model.

\section{The definition of the order parameter in mean field theory}
\label{app:Q}
The construction of the order parameter is quite complex~\cite{ParisiQ,MPV}: we consider an
infinite number of equilibrium configurations $\{\sigma^\alpha\}$, with
$\alpha=1,\infty$. We can compute the overlaps
$q^{\alpha,\gamma}\equiv\frac{1}{L^D}\sum_i \sigma^\alpha_i \sigma^\gamma_i$. Let us
denote by $\mathcal{Q}$ this infinite matrix. The probability distribution of
the values $Q$ in the overlap matrix $\mathcal{Q}$ defines a function $P_J(Q)$
which changes with the disorder configuration $\{J\}$. Denoting by
$\left[\cdots\right]_J$ the average over different instances,
we finally define $P(Q)=\left[P_J(Q)\right]_J$ as the average over $\{J\}$ of
$P_J(Q)$. The function $P(Q)$ is the order parameter of spin glasses in the
mean field limit. Indeed it is possible to construct a functional free energy
$F[P]$ such that in the mean field limit one can obtain the equilibrium value of
the free energy by a variational principle. All this has been rigorously
proven~\cite{Talagrand} in the Sherrington-Kirkpatrick~\cite{SK} model where the mean
field theory was supposed to be correct.

\section{The computation of the free energy barrier as function of $\text{\textit\bf{M/L}}$}
\label{app:var}
Let's consider a model as simple as a chain of $L$ continuous variables $q_{i}$
with periodic boundary conditions, for which we write the analogous of the
free energy cost (see eq.~\ref{eq:Ftheta}) as
\begin{align}
\label{eq:min}
\Delta F[q]=\sum_i|q_i-q_{i+1}|^{5/2}- \sum' q_i + \sum''q_i\;,
\end{align}
where the second and third terms are sums over over two regions of $M\leq L/2$
consecutive sites, taken at the largest possible distance $L/2-M$. We may take
$\sum'$ in the region $0\leq i<M/2,\; L-M/2\leq i<L$, and $\sum''$ in the
region $L/2-M/2\leq i < L/2+M/2$.
%% Along with PBC symmetry
%% ($q_{i}=q_{i+L}$), we also impose symmetry under reflections about $i=L/2$
%% ($q_{i}=q_{L-i}$), and anti-symmetry under reflections about $i=L/4$ and
%% $i=3L/4$ ($(q_{i}=-q_{L/2-i}$, then $q_{L/4}=q_{3L/4}=0$).
Such terms favor configurations with opposite total overlaps
in the two regions of size $M$.
We can numerically minimize the action above for any given $M$, finding an optimum $q^o_i$ and computing
$\Delta_M=(1/M)\left|\sum'q^o_i\right|=(1/M)\left|\sum''q^o_i\right|$
and then the quotients $A_{M,L}=\Delta F[q^o]/\Delta_M^{5/2}$ as functions of
$M$ and $L$

We can also do an analytic computation directly in the continuum limit considering
the functional:
\begin{align}
\label{eq:Fcont}
\Delta F[\partial_xq(x)]=\int_0^1 dx\left|\frac{\partial q(x)}{\partial x}\right|^{5/2}\; ,
\end{align}
with $q(x)$ a function in $[0,1]$ with the following constraints and symmetries: we impose
$q(0)=q(1)=q(1/2)=0$, $q(x)=-q(1-x)$, $q(x)=q(1/2-x)$ for $x<1/2$ and then $q(1/4)=Q_0$,
$q(3/4)=-Q_0$ with, say, $Q_0>0$. We consider the
variational problem with the constraint that the sum of the total
overlaps in two regions of size $z$, one centered in $x=1/4$ and the other in
$x=3/4$, with $0<z <1/2 $, must be zero:
\begin{align}
\Delta F_{\lambda}[q,\partial_xq(x)] &=\int_0^{1} dx\left|\frac{\partial
  q(x)}{\partial x}\right|^{5/2} \\
&-\lambda \left[\int_{\frac{1}{4}-\frac{z}{2}}^{\frac{1}{4}+\frac{z}{2}}dxq(x)+\int_{\frac{3}{4}-\frac{z}{2}}^{\frac{3}{4}+\frac{z}{2}}dxq(x)\right]\;,\nonumber
\end{align}
Given the symmetries, the extremum and the dependence of the multiplier
$\lambda$ on $z$ are obtained requiring
regularity of the extremal function $q(x;\lambda)$ and its derivative in $x=1/4$
and $x=1/4-z$. The expression derived for $q(x;\lambda)$ is used for computing
the free energy cost $\Delta F[q(x;\lambda)]$ (equation~\ref{eq:Fcont}),
the overlap fluctuation $\Delta(z)=(2/z)\int_{1/4-z/2}^{1/4}q(x;\lambda)$ and,
up to an overall factor, the ratio
$A(z)=\Delta F/\Delta(z)^{5/2}$ as a function of $z$, which plays the role of
$M/L$ above.

The two computations give similar results. Figure~3
shows the results of the analytic computation in the continuum limit
compared to the coefficients $A_{M,L}$ appearing in
equation~\ref{eq:Pdelta2}, and obtained by analyzing Monte
Carlo data. As one can see, barring renormalization constants, values of $A_{M/L}/A_{M'/L}$
obtained from Monte Carlo data compares well to estimates by these two methods.

\begin{figure}
\centering
\includegraphics[width=0.9\linewidth]{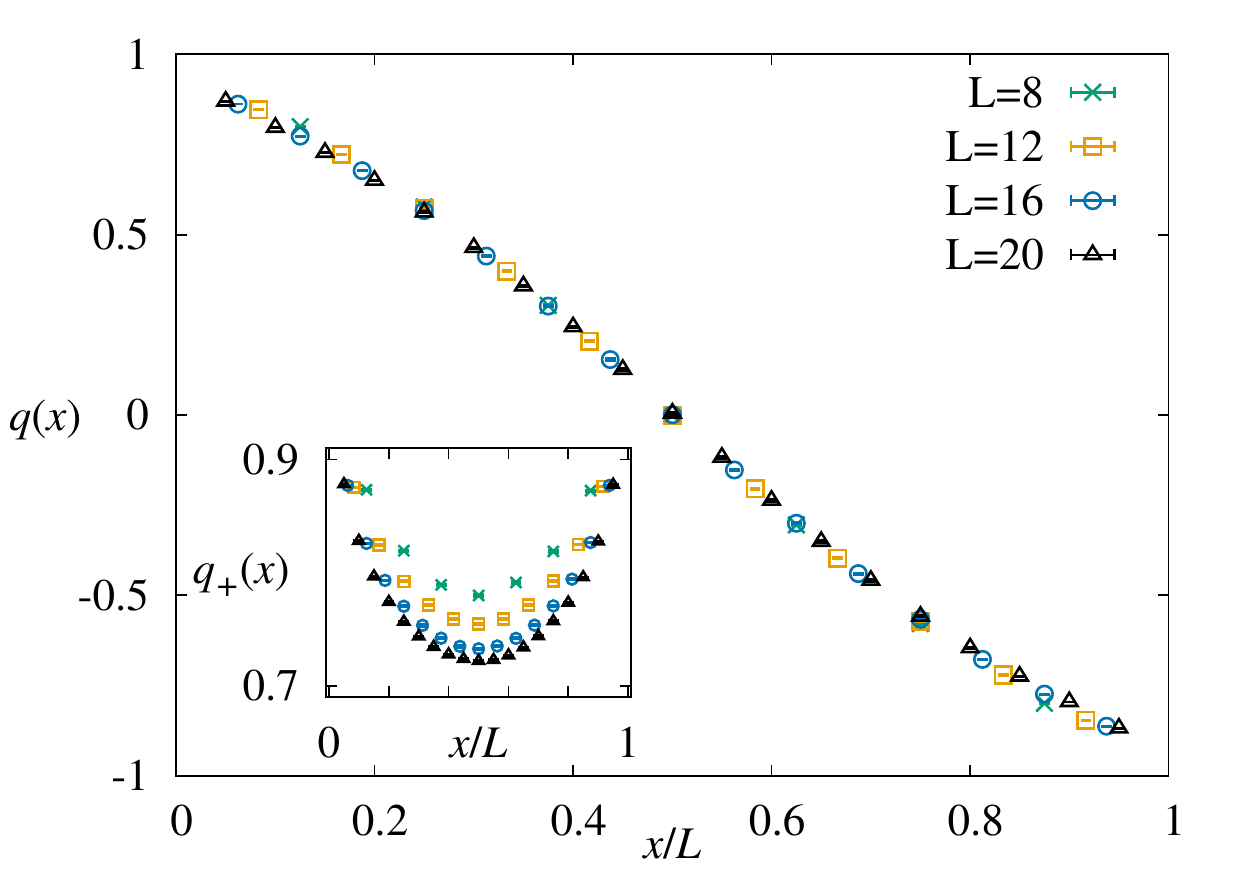}
\caption{The overlap $q(x)$ between the $S^{+}$ (PBC) and $S^{-}$ systems (see
section \emph{Methods}, equation~\ref{eq:qx}, and the appendix~\ref{app:cfr}), as a function of $x/L$. In the inset, the overlap $q_+$between $S^{+}$
and the reference configuration of the non-constrained system, from which $S^{+}$ and
$S^{-}$ have been replicated, as a function of $x/L$.}
\label{fig:qx}
\end{figure}

\section{Comparison to previous estimates of the stiffness exponent}
\label{app:cfr}
The  overlap protocol used in the simulations here yields a different scaling
from domain-wall stiffness exponent considered in \cite{Boettcher1,
  Boettcher2, Boettcher3, Hartmann}, although both analyses agree on the value
of the lower critical dimension. Let us summarize the main different features
of the protocol: 
\begin{enumerate}
\item Our approach involves a thermal and a disorder average at finite temperature instead of the pure
disorder average  at $T=0$ in \cite{Boettcher1, Boettcher2, Boettcher3, Hartmann}.
\item   We go from periodic to antiperiodic boundary conditions as in
  \cite{Boettcher1, Boettcher2, Boettcher3, Hartmann}. However, we add the
  further constraint that the spins of a hyperplane remain fixed. 
\end{enumerate}
We believe that condition (2) makes the main difference while (1) is a just a
technical difference: however it would be extremely interesting to check
numerically what happens a $T=0$ on the same samples following both
protocols. 

The previous investigations~\cite{Boettcher1, Boettcher2, Boettcher3, Hartmann} on
scaling with size of domain-wall energy in the EA model (equation~\ref{eq:H}) considered energy
differences $\Delta E_J$ of ground states of a spin glass sample (a given disorder
configuration) upon changing from periodic (PBC) to antiperiodic boundary
conditions (APBC) in one direction.

In ground state computations, random interfaces induced by the
change in boundary conditions adjust to and take advantage of bond
frustration to minimize their energy. Given the random disorder, the sample
average of the energy differences $\Delta  E_J $ between PBC and APBC
vanishes. The relevant quantity is the absolute value $|\Delta E_J|$; on the contrary in our case the energy variation is always positive definite.
One expects for the sample average of the magnitude of the difference to scale
as $\left[|\Delta E_J|\right]_J\sim L^{\mathcal{Y}_D}$
with the \emph{stiffness} exponent $\mathcal{Y}_D>0$ at dimensions above $D_{lcd}$ and $\mathcal{Y}_{D_{lcd}}=0$.
The estimates~\cite{Boettcher2} for the $D$ dependent stiffness exponent are
$\mathcal{Y}_{D=3}\simeq0.24$ and $\mathcal{Y}_{D=4}\simeq 0.61$, which are well
below the estimates for interface energy exponents in the present
work (our exponents are roughly speaking a factor two larger). Of course, we are studying our systems at finite temperature, and we cannot
exclude a dependence on $T$ of the stiffness exponent, but we are actually
considering the scaling properties of quantities which are distinct from
(albeit probably non-trivially related to) the ones considered in ground state studies.

In order to test the FPV prediction (equation~\ref{eq:FPV})
we need to constrain two independent replicas $S^{+}$ and $S^{-}$ for a given sample as described in the
\emph{Methods} section. The two replicas have a given total mutual
overlap fluctuating around $Q=0$ (in our specific case) and the maximum possible overlap
difference along one direction $\Delta=|q(x=0)-q(x=L-1)|$, where $q(x)$ (see
equation~\ref{eq:qx}) is the overlap on the plane orthogonal to the $x$
direction and of given $x$ coordinate. Freezing all spins on the $x=0$ plane
(to values of an equilibrium configuration of the original system)
on both replicas ($q(x=0)=1$) and imposing APBC along the $x$ axis on $S^{-}$ are
the device by means of which we impose the constraint. The frozen spins act as
an external field breaking the underlying symmetries and favoring configurations with
large positive mutual overlap in the system $S^{+}$ which retains periodic boundary
conditions. In $S^{-}$, with APBC, an interface develops as a result of the energy cost
introduced by flipping the boundary. Since the spins on the border are frozen
and drive the system towards configurations strongly correlated to the
equilibrium configuration of $S^{+}$, the configuration of spins in $S^{-}$ is
not free to relax toward an equilibrium state which would be typical of its
disorder configuration (factoring in APBC), resulting in a positive average energy difference
$\Delta E$ at the interface between two spin-reversed phases, as in the case
of the Ising ferromagnet.

In addition, given the frozen configuration on the
$x=0$ plane, the interface cannot cross the boundary
without additional free energy cost; entropic repulsion pushes it towards the
central region (it moves away from the wall in
the search of space to fluctuate)~\cite{Interfaces}.
As a result, the induced interface is not as free to
adjust to frustrated links as in the standard ground state
computations, giving rise to a different definition of the scaling exponent.

The situation would be analogous, although more complicated, in the general
case, say, $Q>0$ and $\Delta < 2-2Q$ which
is harder to implement in a direct simulation. In that case the interface
would result from the competition of two phases with overlaps $Q-\Delta$ and
$Q+\Delta$ at opposite borders, not related by any simple symmetry.

We show the overlap $q(x)$, equation~\ref{eq:qx}, between systems $S^{(+)}$ and
$S^{(-)}$ as a function of $x/L$ in figure~\ref{fig:qx} for $D=3$, $T=0.7$, in
our case of study: $Q>0$ and $\Delta = 2$. This scaling shows that we are not
far from the asymptotic limit.
In the same figure we also show, as a function of $x/L$, the overlap
$q_+(x)=\frac{1}{L^{D-1}}\sum_{i_1,\dots,i_{D-1}}\sigma_{i_D=x}^{(\mbox{ref})}\sigma_{i_D=x}^{(+)} $,
where $\left\{ \sigma^{(\mbox{ref})}\right\}$ is the equilibrium spin
configuration of the original system from
which $S^{+}$ and $S^{-}$ are replicated (see section \emph{Methods}). $q_+(x)$ takes large values for all
$x$, well above the value of $q_{EA}(T=0.7) \simeq0.52$.~\cite{JanusEq,JanusQ}

We have observed that in the case of PBC-APBC, the energy difference $\Delta
E$ has a random sign: the average of $\Delta E$ is zero  and the interesting
observable is the average of $|\Delta E|$. In our case the energy variation
$\Delta E$  has a non zero average;  in three and four dimensions  this
average is asymptotically larger than the average of  $|\Delta E|$ with the
other protocol (the exponent is nearly a factor two larger).

We would like to further speculate on the possible physical origins of this
difference. Let us discuss what happens in the PBC-APBC protocol. Changing the
boundary condition at $T=0$ introduces a relative interface plane between
spins that are flipped and the spins that are not flipped. A crucial point is
the fractal dimensions of this interface. Two different scenarios are possible
\begin{itemize}
\item In the large volume limit, this interface is essentially localized around a
plane dividing the system into two regions: a region where the new ground
state is equal to the original ground state and a region where the new
ground state is the spin reversed one. The interface could be rough, but as
far as its fractal dimension $d_F$ is less than the space
dimensions $D$ the value of $d_F$ is irrelevant.  In this scenario, the
interface of our work should behave in a way quite similar to the interface
of PBC-APBC . Indeed, for large volume, the interface we create  should  be
able to avoid the parallel plane of fixed spins, since its location along
the $D$-direction is arbitrary. In this scenario, the energetics of this
protocol should be the same of PBC-APBC.
\item  In an alternative scenario in the large volume limit this interface is
not localized in a particular region and it is space filling
($d_F=D$). There are no large regions where the new ground state is similar
to the old ground state. In this scenario, if we force the system in a given
region to be similar to the original one, we have to   pay an additional
energy cost and this explains the difference between the two protocols.
\end{itemize}
The first scenario has been advocated by Wang et al.~\cite{WWANG} in the
framework of an approximate computation of the ground state;  the second
scenario has been advocated by Marinari and Parisi~\cite{MPinterf} using
information from exact ground states on systems of  sizes up to $14^3$.

We think that it would be very interesting to use modern technologies to
check which of those two scenarios is the correct one at $T=0$, also in order
to clarify the origin of the difference of our exponents with those of
\cite{Boettcher1, Boettcher2, Boettcher3, Hartmann}.

\end{document}